# BIOPHOTONS - NEW EXPERIMENTAL DATA AND ANALYSIS

M. Benfatto[1], E. Pace[1], I. Davoli[2], Massimiliano Lucci[2], R. Francini[3], Fabio De Matteis[3], Alessandro Scordo[1], Alberto Clozza[1], Maurizio Grandi4, C. Curceanu[1] and P. Grigolini[5]

[1] *Laboratori Nazionali di Frascati - Istituto Nazionale di Fisica Nucleare - Via E. Fermi 40 - 00044 Frascati (Italy);* maurizio.benfatto@lnf.infn.it, elisabetta.pace@lnf.infn.it, catalina.curceanu@lnf.infn.it
[2] *Dipartimento di Fisica - Università di "Tor Vergata" - Via della Ricerca Scientifica - 00133 Roma (Italy);* ivan.davoli@roma2.infn.it
[3] *Dipartimento di Ingegneria Industriale - Università di "Tor Vergata" - Via del Politecnico - 00133 Roma (Italy);* roberto.francini@roma2.infn.it
4 *Istituto La Torre - Via M. Ponzio 10 - 10141 Torino (Italy);* latorre@mauriziograndi.it
[5] *Center for Nonlinear Science, University of North Texas, Denton, Texas (USA);*
paolo.grigolini@unt.edu

Biophotons are an ultra-weak emission of photons in the visible energy range from living matter. In this work we study the emission from germinating seeds using an experimental technique designed to detect light of extremely small intensity. The emission from lentil seeds and single bean was analyzed during the whole germination process in terms of both the different spectral components through low pass filters and the different count distributions in the various stages of the germination process. Although the shape of the emission spectrum appears to be very similar in the two samples used in our experiment, our analysis is able to highlight the differences present in the two cases. In this way it was possible to correlate the various types of emissions to the degree of development of the seed during germination.

1. **Introduction**

   Nearly an hundred years ago, the Russian biologist A. Gurwitsch, [1,2] doing experiments with onion plants by measuring their growth rate, observed that this was strongly influenced by the fact that the various seedlings were close or not and that this behavior remained even if any possibility of bio-chemical exchange had been eliminated. On this basis he hypothesized the presence of a weak electromagnetic field emitted by the plants that somehow is the responsible for the regulation of cell growth. This interesting observation was completely forgotten by the scientific community for many years and only in the 1950s an electromagnetic emission from living organisms was highlighted with the work of Colli and Facchini [3,4]. Later, in the 80s, F.A. Popp [5] did an extensive work in this field to understand in details the origin and the meaning of such ultra-weak emission, hereby called bio-photons. They are an endogenous production of ultra-weak photon emission in and from cells and organisms, and this emission is characteristic of alive organisms. This emission is completely different from the normal bioluminescence observed in some simple as well complex organisms, it is at least 1000 times weaker and it is present in all living organism, from plants to human beings. It is also different from the thermal radiation because its intensity is many orders of magnitude higher than the one calculated by Planck's law in the visible energy range at room temperature. The fact that this emission ends when the organism dies completely excludes the possibility that it is the product of either some radiative decay produced by traces of radioactive substances present in the organism or by the passage of cosmic rays.

   The main characteristic of biophotons [5,6] is a total intensity of the order of one hundred photons/sec with a practically flat emission within the energy range between 200 and 800 nm. After any type of stress due, for example, to some chemical agents or excitation by light, the emission increases of almost a factor ten and relaxes to the normal values quite slowly, following hyperbolic functions rather than an exponential law.

   Despite the wealth of experimental results, the questions of what biophotons are, how they are generated and how they are involved with life are still open. There are two hypotheses about it [5,6]. The first sees the emission as the random radiative decay of some molecules excited by metabolic events while the second hypothesis assigns the emission to a coherent electromagnetic field generated within and between the cells by some biochemical reactions in which, perhaps, oxygen atoms are involved.

   At the same time, there are several experimental evidences that such radiation carries important biological information [7,8], for example, the radiation emitted by growing plants or organisms can increase the cell division rate in similar organisms by as much as 30%, the so-called mitogenetic effect [9,10,11].

   Typically our group deals with bio-photon emission from the germination of seeds of various kinds. The experimental set-up to measure the biophoton emission of germinating seeds is based on photomultiplier techniques and it is constituted by a dark chamber and a photomultiplier sensitive to the visible energy range. The detector works as a photon counter and the data are recorded as the number of photons detected, within a well-defined and adjustable time window of measurement, as a function of time. The measurements are taken during periods varying from few hours to several days, depending on the sample. For this reason, the typical photon-counting data set is a time series where the number of counted photons is reported in the ordinate axis as function of time measured starting from a time zero which can be chosen at will, and typically is the moment in which the measurement is started after having closed the dark chamber. Details can be found in the next section and ref. 12.

In a recent work our group proposed the Diffusion Entropy Analysis (DEA) approach to analyze the time series produced by the photon counting of germinating lentils. The method of diffusion entropy analysis was introduced in 2001 in ref. [13] and is based on the diffusion approach to evaluate the Kolmogorov complexity. The Kolmogorov complexity is turned into a scaling η, that is expected to depart from the ordinary value η = 0.5. The experimental time series, like the emission we record with our experimental set-up, is converted into a diffusional trajectory. The complexity of the signal is derived through the evaluation of the Shannon entropy associated to the diffusional trajectory [12,13,14]. The main result of ref. 11 is that the biophoton emission shows conditions of anomalous diffusion with a substantial deviation of the scaling coefficient from the ordinary value η = 0.5. Always in this work it is highlighted as at the beginning of germination the condition of anomalous diffusion in realized with the clear presence of crucial events. On the other side, when the seeds generate roots the complexity of the biophoton signal changes completely its nature and the departure from the condition of random diffusion is due to Fractional Brownian Motion (FBM) [15]. This result yields an impressive similarity with which found by the authors of Ref. 16, who analyzed the heartbeats of patients under the influence of autonomic neuropathy. In this case the increasing severity of this disease has the effect of moving from a complexity condition generated by crucial events to a complexity condition characterized by the FBM infinite memory.

In the bio-photon emission of germinating lentils, the transition from a complexity characterized by crucial events to a condition dominated by FBM could indicate the transition from a normal physiological condition to one of stress due to the lack of light in the experimental chamber but also the fact that the behavior of plants during the growth process is completely different from that of human beings [11,12]. Plants do not have well-defined organs, only at the beginning of the germination process there is a differentiation process that may require the presence of crucial events. In other words, we cannot rule out that the kind of criticality involved by the germination process requires a form of phase transition that is not yet known. It has to be stressed that Mancuso [17] and Mancuso and Viola [18] use the concept of swarm intelligence with reference to the non-hierarchical root network. Thus, it may be beneficial to supplement their observations noting that the initial region of germination may have to do with the birth of this surprising root intelligence.

In this work we present a new analysis of the bio-photon emission in terms of both the various spectral components and of the photo-counting distribution functions in various conditions both for lentil seeds and for the new experimental data relating to a single bean.

## 2. Experimental data and comparison

Our experimental set-up is formed by a germination chamber and a photon counting system. Seeds are kept in a humid cotton bed put on a Petri plate. The photon counting system consists of an Hamamatsu (H12386 110) counting head placed on top of the germination chamber and an ARDUINO board driven by a PC with Lab-View program. The acquisition time window is fixed at 1 second. The germination chamber is built with black PVC to avoid any contamination of the light from outside. The whole system has a dark current of about 2 photon/sec at room temperature. Without any seeds or germination there is a monotonic decrease of photon emission which arrives in few hours to the value of the electronic noise. This emission tail comes from the residual luminescence of the materials, consequence of the light exposure of the experimental chamber.

The experiment was done using lentil seeds (76 seeds) and a single bean seed. The results are shown in figure 1. Panel a) shows the emission of the 76 lentils while panel b) shows the signal coming from the single bean.

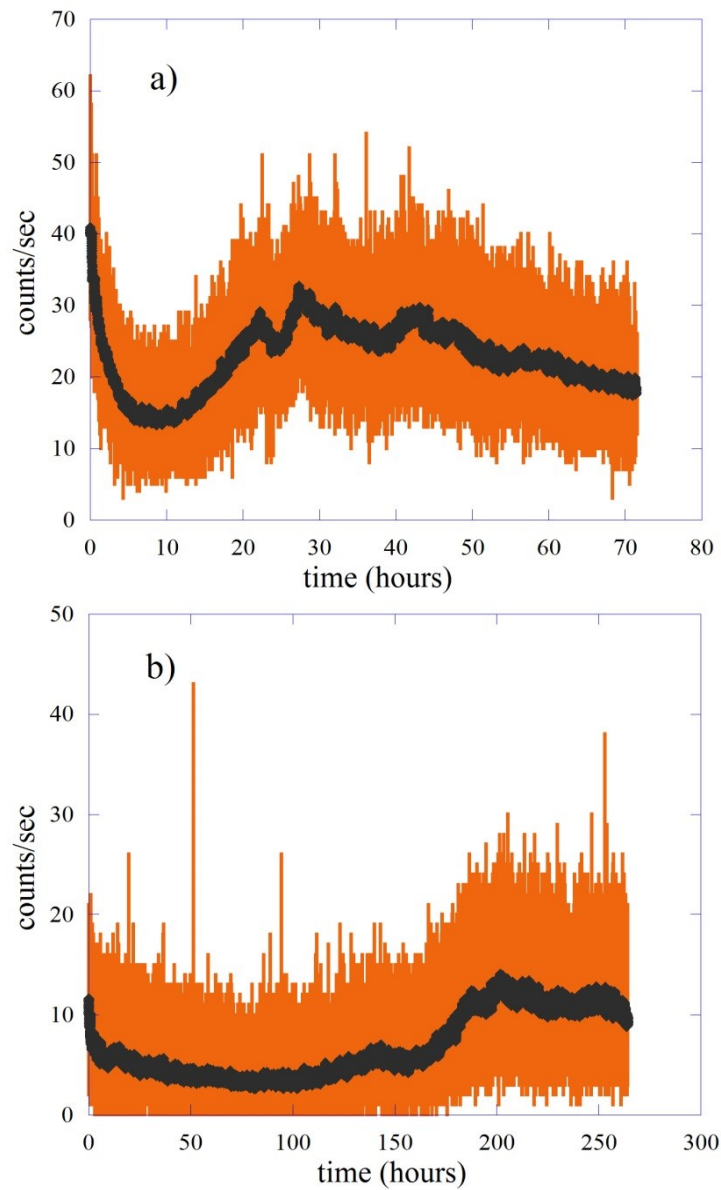

Figure1 – biophoton emission of the lentil seeds (panel a) and single bean (panel b). The black curves are the counts per second averaged over 1 minutes.

In both cases the emission is characterized by the initial tail of residual luminosity which ends in a few hours and then by a rise which indicates the beginning of germination and the subsequent development of the roots and then of the leaves. Note how the time scale is completely different in the two cases.

In figure 2 we present the comparison between the two emissions. For clarity, only a single time scale has been used and the figure shows the emissions in the unit of the counts per second averaged over 1 minute. To use a single time scale we chose to align the two maximums of the emission (peaks C), this led to multiply the time scale for the single bean by the factor 0.164. The two curves have been moved further to have the zero of the time scale positioned in the first minimum. This means that the value 10 and the value 100 respectively for the lentils and for the single bean have been subtracted from the original time scale.

To have the same number of counts in peaks C the values of the counts relating to the single bean were multiplied by a factor 2.28.

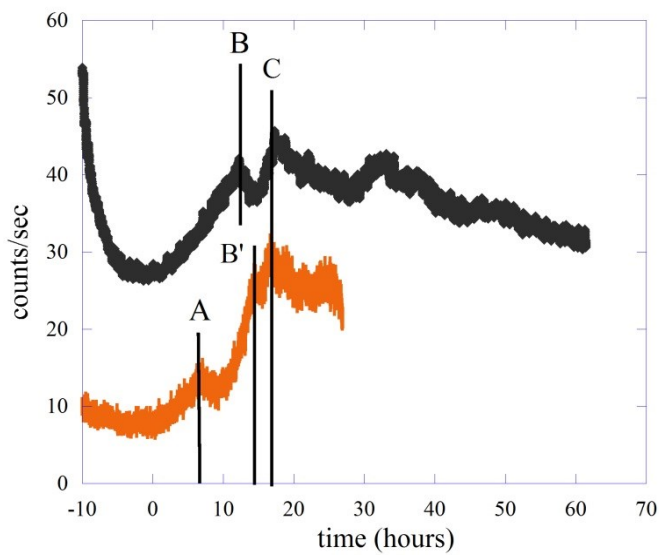

Figure 2 – Comparison between the biophoton emission of the single bean (orange curve) with the emission of the 76 seeds of lentil. For clarity, the curve relating to the emission of lentils has been moved upwards and it has been used the time scale of the lentils emission. To get the time scale of the single bean multiply the numbers by the factor 6.1 and add 100 hours.

It is interesting to note that the two types of seeds have a very similar form of emission. In the time phase between 0 and 20 the lentil emission presents two peaks (B and C) separated by about 5 hours, the same two peaks (B 'and C) are present in the emission of the single bean but here they are separated by about 14 hours. The biophotonics emission of the bean shows a further peak (peak A) about 43 hours after the minimum positioned at zero in the scale. This peak is absent in the lentil emission. It should also be noted that in the germination phase, between zero and peak C, the growth phase of the emission presents at least two slopes.

This type of temporal evolution of the light emission seems to be a general behavior of biophotons coming from seeds in the germination phase, emissions of common wheat [19] (Triticum aestivum) and of seeds of Arabidopsis thaliana [20] have very similar behavior of what presented here in figures 1 and 2.

By renormalizing the emissions by the number of seeds, we can find the ratio between the number of photons emitted for each single seed of different type. In this experiment the ratio between the number of photon emitted by the lentils and the number of photon emitted by the single bean is about $1/33$, number very close to the ratio between the averaged weight of single seed [21].

### 3. Data analysis I – Probability distribution functions

In this paragraph, we report the analysis of the emitted light in terms of the probability distribution function $P_m(T)$ of finding m counts in a given acquisition time window T. In a semiclassical picture of the optical detection process the phototube converts the continuous cycle-averaged classical intensity $\bar{I}(t)$ in a series of discrete photocounts. Thus the number m of photocount obtained in an integration time T is proportional to the intensity of the light that arrives on the detector [22].

A photocount experiment consists of a sufficiently large number of measurements of the number of photocounts in the same integration period T. The result of the measurement is expressed by the function $P_m(T)$ which represents the probability of obtaining m counts in the acquisition time T. The purpose of this measurement is to determine, if possible, the statistical properties of light through the properties of the distribution function, considering that, at least in some particular cases, there is a direct correspondence between the functional form of the $P_m(T)$ and the statistical properties we are looking for.

In a real experiment it is practically impossible to repeat the same experiment many times, in particular in our case where we have a system that is germinating and which therefore changes over time although with much longer times than acquisition time window T, which is $1\ sec$ in our experiment. The distribution function is thus determined by means of a series of observations of duration T, i.e. we detect the number of photons arriving in the phototube in one second and do this for the whole duration of the experiment, typically from one hour to the whole time interval of the data set.

The required photocounts distribution function is obtained as an average over successive starting time t of the function

$$P_m(t,T) = \frac{[\xi \bar{I}(t,T)T]^m}{m!} \exp[-\xi \bar{I}(t,T)T] \qquad (1)$$

where $\xi$ is the detector efficiency and $\bar{I}(t,T)$ is the mean intensity of the light field on the phototube in the period from $t$ to $t+T$ [22]. So $P_m(T) = \langle P_m(t,T) \rangle$ and the average is done as previously described. On this basis, the mean number of counting is easily obtained as $\langle m \rangle = \sum_m m\, P_m(T)$ as well the different moments and the variance of the distribution. It has been assumed that the emission is stationary. In our case this is not strictly true but this assumption becomes a good approximation for time intervals of the order of an hour or especially after the growth phase at the beginning of germination.

There are only some special cases where the average can be done in an analytical form. The simplest is that of a stable classic light wave where $\bar{I}(t,T) = \bar{I}$, i.e. the cycle-averaged intensity has a fixed value independent on the time[22] . In this case the distribution $P_m(T)$ has Poissonian form like

$$P_m(T) = \frac{\langle m \rangle^m}{m!} \exp(-\langle m \rangle) \qquad (2)$$

where $\langle m \rangle = \xi \bar{I} T$. A Poisson distribution is a sign of a system in a coherent state which corresponds to a classical electromagnetic waves [22,23], but, at the same time, this distribution also occurs for experiments where the integration time $T$ is much longer than the characteristic time of the intensity fluctuations of the light beam. For Poisson distribution the variance is equal to the average $\sigma^2 = \langle m \rangle$, any departure from the Poisson distribution is an indication of a non-classical nature of the light and can be measured by the Fano factor [23] $F$ defined as $\sigma^2 = F \langle m \rangle$.

The photocounts distribution can be also derived for a complete chaotic light [22] and it results equal to the photon distribution in a single mode of a thermal source:

$$P_m(T) = \frac{\langle m \rangle^m}{(1+\langle m \rangle)^{1+m}} \qquad (3)$$

and can be applied to any chaotic light of almost any type [22]. This expression can be generalized for thermal source with M modes [23]:

$$P_m(T,M) = \frac{(m+M-1)!}{m!(M-1)!} \left(1 + \frac{M}{\langle m \rangle}\right)^{-m} \left(1 + \frac{\langle m \rangle}{M}\right)^{-M} \qquad (4)$$

Thermal states are classical and there is the relation

$$\sigma^2 = \langle m \rangle + \frac{\langle m \rangle^2}{M} \qquad (5)$$

between the average number of count and the variance. In general the coefficient $M$ can be very large, this means that the variance becomes almost equal to the average value and we find the same relationship valid for the Poisson distribution. As consequence, for very large $M$ (greater than 20) [23] the thermal photocount distribution approaches the Poisson distribution. This implies that it is very difficult to discriminate between coherent and thermal states when many modes are present, in agreement with the discussion of Ref. [23].

We start our analysis from the counts coming from the dark, i.e. the counts measured with the black cap. In other word it is just the noise measured by the Hamamatsu counting head. In Fig.3 we report the comparison between the experimental $P_m(T)$ (red square) and two fits using equations 2 (blue line) and equation 3 (green line) respectively. The experimental average value is $\langle m \rangle = 1.56$ and the variance is $\sigma^2 = 2.24$ [12], the distribution turns out to be of super Poissonian type with a Fano factor $F = 1.43$, slight bigger than one. The two fits are almost equivalent in terms of agreement with the experimental data having an almost equal $\chi^2$ [24], although they produce a different value of the average counts, $\langle m \rangle = 0.827 \pm 0.03$ for the Poisson case and $\langle m \rangle = 1.16 \pm 0.17$ for the other case. This last value is much more in agreement with the experimental $\langle m \rangle$ and both values are consistent with the dark count data of this phototube [25].

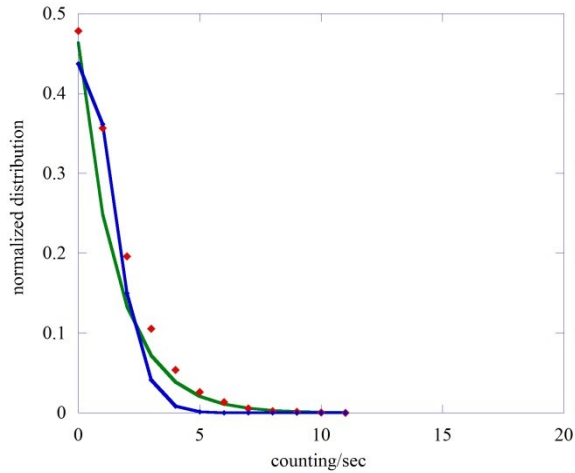

Figure 3 - Comparison between the experimental counting distribution function (red squares) and two fit performed with a Poisson function (blue line) and a single mode thermal state distribution function (green line).

We then proceeded with the same type of analysis of the emission in the presence of seeds. In the reference [12] the count probability distribution functions $P_m(T)$ for the case of lentils have been derived using 1 hour of emission at different germination time. The result of that analysis was to have a slight tendency to super Poisson distributions in all cases, this being typical of chaotic or partially coherent sources.

In this paper we present a similar analysis but using different size of the emission period, up to the entire measured data. The analysis relating to the single bean will also be presented.

In Fig.4 we show the comparison between the experimental $P_m(T)$ for the 76 lentils (panel a) and the single bean (panel b) with different type of fits. The measurement time period used to obtain the count probability distribution function is from time 10 (hours) and time 83 (hours) to the end for the lentils and the single bean respectively. See Fig.1 for clarity. In other words, we did not consider in both cases the first period where we could have contaminations due to the residual luminescence, and we waited for the first hints of rising in the counts that indicated (especially in the single bean) the beginning of the germination process. The experimental average value of the counts is $\langle m \rangle = 23.5$ in the case of lentils while the single bean gives a value $\langle m \rangle = 7.9$, the variance is equal to 39.9 and 20.2 for lentils and single bean respectively. In both cases this value is much bigger than the average count.

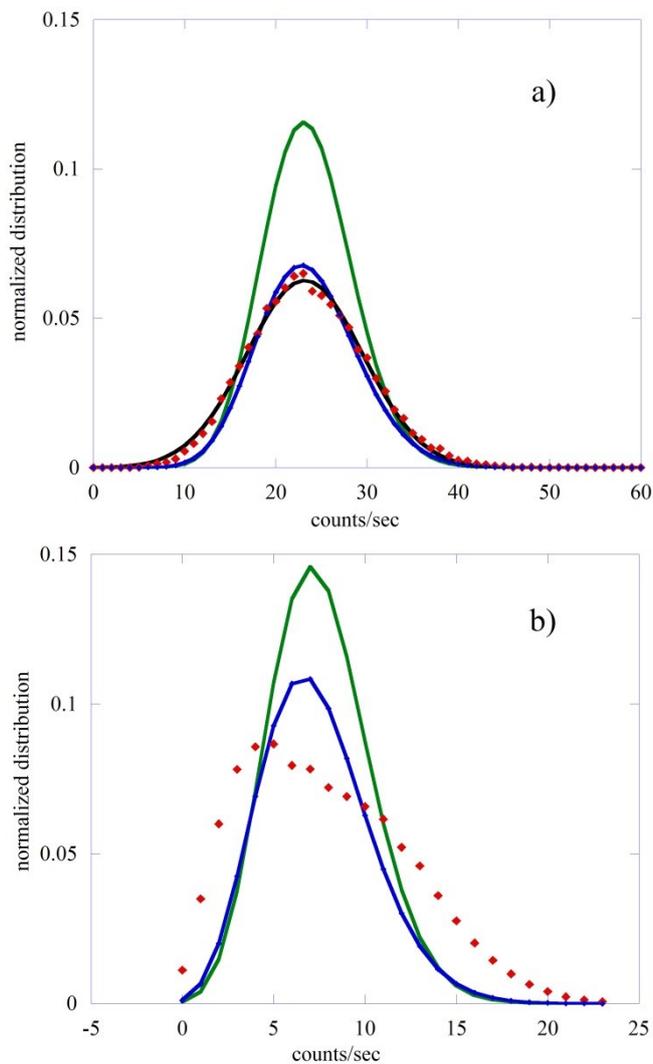

Figure 4 - .Comparison between the experimental count probability distribution function (red squares) of the lentil seeds (panel a) and the single bean (panel b) with three type of fits. The green line refers to a Poisson distribution function, the blue line to the multi-mode distribution function and the black line in panel a) to a Gaussian distribution function.

It is interesting to note how the distribution relative to the single bean is strongly asymmetric, and it is impossible to have a good fit of it using both a Poissonian functional form and a multi-modal thermal type. On the contrary, the distribution relative to lentils can be optimally fitted with both a Gaussian function and a multi-modal thermal function. In both cases the fit with a Poisson function is very bad, but this is not surprising considering the experimental difference between the mean values and variances and that the stationarity hypothesis is only weakly satisfied having considered the entire time interval of measurement.

The fact that the distribution of the 76 lentil seeds is strongly symmetrical and can be optimally fitted with a Gaussian is a clear indication that the various seeds have different germination times which therefore give rise to emissions that are not in phase with each other.

The use of a shorter time period for the calculation of the probability distribution function makes the hypothesis of stationarity more easily satisfied. In the case of a single bean data, using a shorter emission period for the calculation of the function $P_m(T)$, we observe a transition to more symmetrical distributions. As an example, in Fig. 5 we report the comparison between the experimental count probability distribution function (red squares), relative to the period from 200 (hours) to the end, with two fits using a Poissonian fit (green line) and a multi-mode thermal function (blue line).

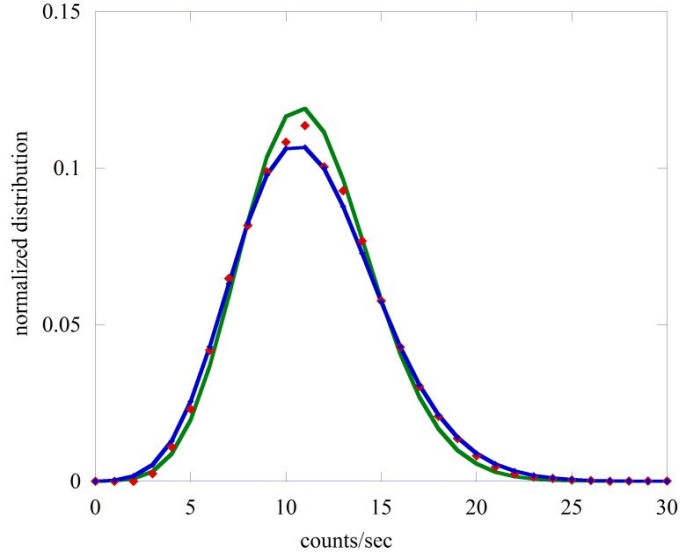

Figure 5 – Comparison between the experimental count probability distribution function (red squares) relative to the single bean emission with two fit using a Poisson function (green line) and a multi-mode thermal function (blue line). Here the emission period is between 200 (hours) and the end of the experiment.

In this case the experimental average counts is $\langle m \rangle = 11.33$ and the variance is $\sigma^2 = 12.93$, value much closer to the average number of count than that found in the previous case. The quality of the two fits is totally equivalent producing a practically identical $\chi^2$ value, in the Poissonian case we obtain a value of the average counts equal to $\langle m \rangle = 11.24 \pm 0.05$, while the multi-mode thermal function gives $\langle m \rangle = 11.3 \pm 0.03$ and $M = 50.0 \pm 0.03$. It is interesting to note that in this last case the eq. 5 is almost satisfied.

We performed this type of analysis for both the single bean and the 76 lentil seeds using emission periods ranging from one hour to several hours, up to the total, as shown previously.. Some of this work is summarized in table 1 where the distribution asym-metry $S$ index is reported only for some of the time intervals considered in the analysis. The asymmetry index $S$ is defined as

$$S = \frac{\mu_3}{\sigma^3} \qquad (6)$$

where $\mu_3$ is the central moment of order 3 and $\sigma$ is the standard deviation. A perfect symmetrical distribution has the value $S = 0$.

| Single bean | | Lentil seeds | |
| --- | --- | --- | --- |
| Time interval | S index | Time interval | S index |
| 82-265 | 0.50 | 10-70 | 0.23 |
| 82-150 | 0.81 | 20-70 | 0.24 |
| 150-200 | 0.30 | 50-70 | 0.28 |
| 200-265 | 0.43 | 35-36 | 0.11 |

Table-1 S values for different time intervals for lentils seeds and single bean. The time interval are measured in hours. Refer to Fig.1 for details.

This analysis confirms that the count probability distribution functions related to the lentil seeds are much more symmetrical than those relative to the single bean in almost all the different time intervals. This may be due both to a different characteristic of the seeds but also to the fact that in the case of lentils we used 76 seeds to have a good signal / noise ratio.

When possible, typically for short time window, the distributions can be fitted with either a Poissonian or a multi-mode thermal function, in any cases the experimental variance is always bigger than mean value $\langle m \rangle$, this indicates a super-Poissonian type of behavior that is typical of either thermal emission or an emission with a very short coherence time compared to the time window of the measurement. This makes very difficult to discriminate between coherent and thermal states using the photo counting distribution analysis, in agreement with the discussion of Ref. [23].

### 4. Data analysis II – the different spectral components.

The use of the a turnable wheel holding a few long pass glass color filters makes possible an analysis in terms of the different spectral components of the emission. The wheel with the filters is placed between the germinating seeds and the detector. The wheel has eight positions. Six are used for the color filters, one is empty and the last one is closed with a black cap [12].

The transmission coefficients of our filters and the efficiency of the phototube as a function of the wavelength of light are shown in Fig.6. The transmission coefficients are essentially theta functions positioned at the wavelengths written in the figure, thus only light with wavelengths greater than the cutoff value shown in the figure can pass. The sensitivity of our phototube allows us to see the emission from near ultra-violet to yellow-orange with good sensitivity.

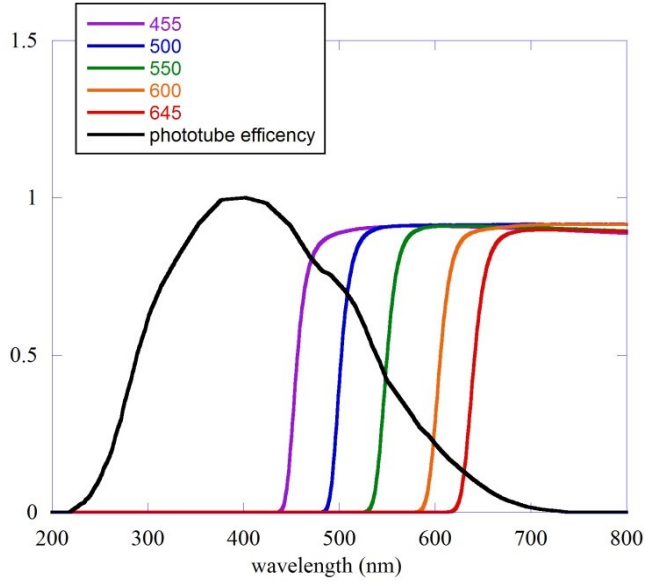

Figure 6 – Transmission coefficients of the different filters used in our experiment and phototube efficiency. The wavelengths of the different cutoffs are reported in the figure.

The total number of counts at time t without filters can be written

$$M_{tot}(t,T) = \int_{\lambda_{min}}^{\lambda_{max}} m(\lambda, t, T)\, \alpha(\lambda)\, d\lambda \qquad (7)$$

where $m(\lambda, t, T)$ is the number of photon emitted from the sample at time t within the integration window of size T at a given wavelength, and $\alpha(\lambda)$ is the efficiency of the phototube. Inserting a filter with a transmission coefficient $f_n(\lambda)$ the number of count $M_n(t,T)$ becomes

$$M_n(t,T) = \int_{\lambda_{min}}^{\lambda_{max}} m(\lambda, t, T)\, f_n(\lambda)\, \alpha(\lambda)\, d\lambda \qquad (8)$$

In Fig.7 we report the quantities $M_n(t,T)$ related to the different filters for both the lentil seeds and the single bean. The counts without any filters is also show for comparison.

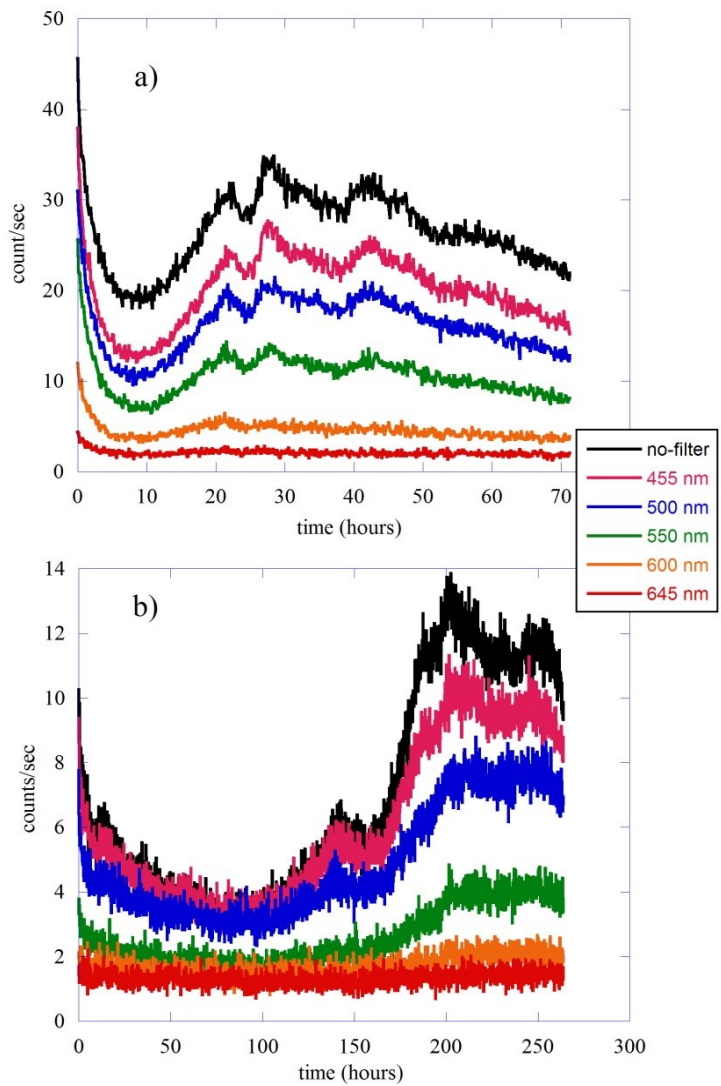

Figure 7 – counts per second averaged over 1 minutes for the different spectral components. Panel a) is related to the lentil seeds while panel b) refers to the single bean.

It is difficult to follow in details the emission with filters, generally speaking we can say that they almost follow the emission without any filters. To see some possible different behavior of the various spectral components we can do a kind of monochromatization calculating the difference between the counts obtained using two filters with adjacent cutoffs, in this way we have almost a monochromatic signal with an energy resolution of the order of 0.1%.

The number of counts detected in the wavelength range defined by two filters with adjacent cutoffs is written as

$$M_{n,s}(t,T) = \int_{\lambda_{min}}^{\lambda_{max}} m(\lambda,t,T)\alpha(\lambda)[f_n(\lambda) - f_s(\lambda)]d\lambda \qquad (9)$$

supposing now that the number of photons emitted from the sample in this wavelength window has a slight dependence on the wavelength, the average number of photon $\bar{m}_{n,s}(t,T)$ in a given wavelength interval can easily derived as

$$M_{n,s}(t,T) \cong \bar{m}_{n,s}(t,T) \cdot \int_{\lambda_{min}}^{\lambda_{max}} \alpha(\lambda)[f_n(\lambda) - f_s(\lambda)]d\lambda = \bar{m}_{n,s}(t,T) \cdot I_{n,s} \qquad (10)$$

$$\bar{m}_{n,s}(t,T) = \frac{M_{n,s}(t,T)}{I_{n,s}} \qquad (11)$$

The value of the $I_{n,s}$ integral can calculate numerically and the average counts can be found on the basis of the differences using Eq.11. In Fig.8 we show the ratio between the different average counts $\bar{m}_{n,s}(t,T)$ and the total signal without filters, for both lentils and single bean. As usual panel a) refers to the lentil seeds while panel b) shows the result for the single bean. Because of the small counts in the case of single bean, the ratio has been smoothed to clarify the behavior as function of time. In the figure the different letters indicate the position in time of the main emission peaks between 0 and the maximum and correspond to the same letters of figure 2.

In both cases the different ratios shown changes as a function of time, in other words, according to the moment of germination, the total signal is formed by different spectral components which change in the relative intensity.

In the case of lentil seeds, the best signal / noise ratio allows us to say that the dominant components are those of orange (600-645 nm) and yellow-green (550-600 nm), in agreement with the results of Colli and Facchini [4]. It is interesting to observe how while the high-energy components remain constant for the entire time of the measurement, the lower-energy parts clearly change in the relative intensity along all the duration of the measurement. In particular in the temporal region between 0 and peak B the orange component is constant at first and then slight decreases, this behavior is associated with a simultaneous increase in the yellow-green component of the spectrum. Here we are in the temporal region where germination began and where DEA analysis tells us that crucial events are present.

The emission curve between 0 and peak C is characterized by having two distinct slopes, one in the time region between 0 and the first maximum (peak B), another between this and the second maximum (peak C). These two temporal regions have different behaviors of the various spectral components of the emission. We have already described the part between 0 and peak B, in the second region (B-C) the yellow-green component is practically constant while the orange component decreases considerably. In other words, the change in slope is associated with a change in the relative weight of the orange and yellow-green spectral components.

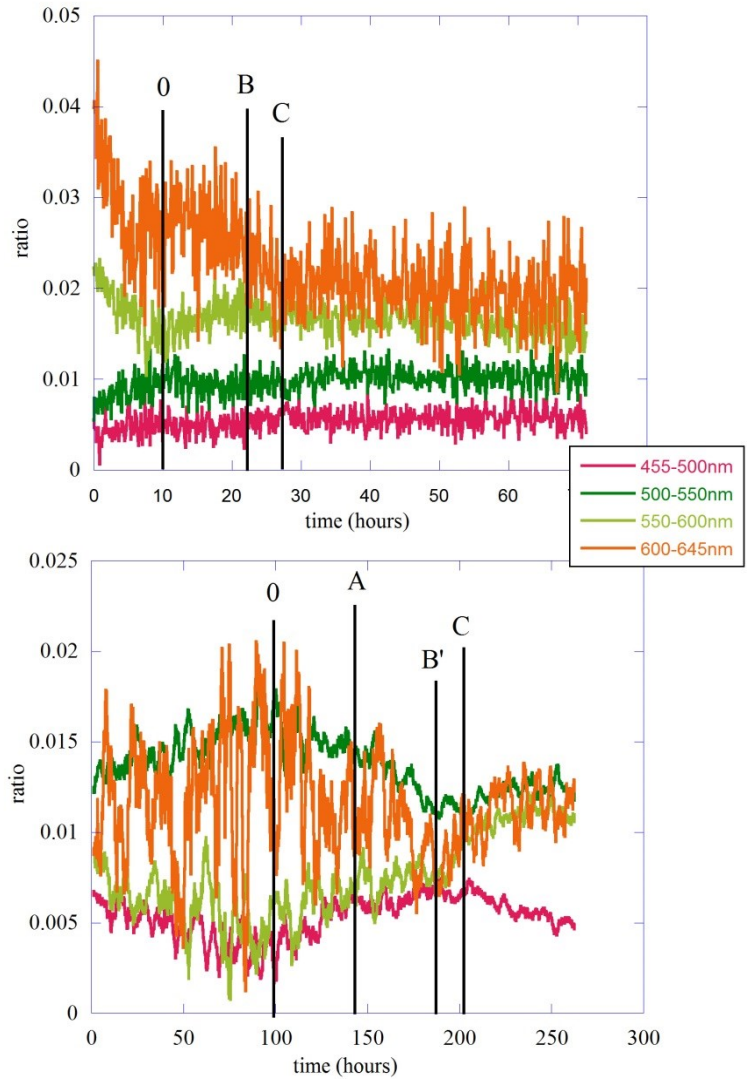

Figure 8 – Ratio between average counts $\bar{m}_{n,s}(t,T)$ and the total signal. The different wavelength windows are reported in the figure with different colors.

The behavior in the case of the single bean is not so obvious because the signal to noise ratio is not good in this case. We can certainly say that in this case the main components are those of orange and blue-green, with the yellow-green component that becomes important in the last phase of germination, where the intensity of emission is at its maximum. In any case, it should be noted that this component grows throughout the duration of the experiment, while the high-energy component (455-500 nm) grows considerably in the region between 0 and peak A and then remains constant in the region between this peak and peak B'. This may possibly be associated with the slope change in the growth of biophoton emission.

5.  **Conclusion and suggestions for future works**

In this work we have analyzed in detail the emission of biophotons from lentil seeds and from a single bean throughout the germination period. We have highlighted the remarkable similarities in the form of emission although the emission associated with the bean is characterized by the presence of an extra peak in the time period delimited by the 0 and the maximum of the emission (peak C), see Fig. 2 and Fig. 8.

Data were analyzed both in terms of the probability distribution functions of counts, using time windows of different sizes, and in terms of the different spectral components of the biophoton emission.

In the case of probability distribution functions, the procedure and the results obtained have been described in detail in paragraph 3. Although this method of analysis has intrinsic difficulties in obtaining reliable information on the statistical properties of the emitted light, it is clear how it should be used mainly in cases where the emission from a single seed is observed to avoid the problems related to the different times of germination of the seeds and which significantly conditions the shape of this distribution as shown in the figure 4 (lower panel).

The analysis of the various spectral components clearly shows how these change throughout the germination period. We have shown how these change their relative weight in the total emissions and how this is in some way connected with the details of the emission, in particular the different slopes observed in the spectra. This behavior could be a clear signal that during the germination period the parts of the seed involved in the emission process change according to the development of the plants.

All this connected must be connected with the information obtainable from the DEA in order to build a credible model for this fascinating natural phenomenon.

It is absolutely essential to increase the signal to noise ratio. This can only be achieved by increasing the solid angle of acceptance of the detector. One possible way is to interpose Fresnel lenses between the sample and the phototube. This method has the advantage of being very cheap. We are carrying out a series of tests to understand the gain in terms of signal.

**Note** - During the preparation period for this work our friend and colleague M. Lucci passed away. It is with great pain and sadness that we report this news. This work also wants to be a way to honor Massimiliano's work and life.

**Acknowledgements**

We warmly thank Giuseppe Papalino and Agostino Raco for their help with the building of the experimental setup. We also thank the conversations and suggestions we had with Dr. I. H. von Herbing, L. Tonello, A. Pease and D. Lambert throughout the duration of the experiment.